\documentclass[11pt]{article}
\usepackage[margin=1in]{geometry}
\usepackage{amsmath}
\usepackage{amssymb}
\usepackage{booktabs}
\usepackage{graphicx}
\usepackage{siunitx}
\usepackage{hyperref}
\usepackage{caption}
\usepackage[section]{placeins} 
\usepackage{everypage}
\usepackage{tikz}
\title{An ensemble of open binary CALPHAD assessments quantifies the uncertainty of the Nd\(_2\)Fe\(_{14}\)B feasibility threshold and surfaces inconsistencies invisible to single-assessment users}
\author{I.~Radulov}
\date{29 November 2026}

\begin{document}
\maketitle

\begin{abstract}
A single merged ternary description built from open binary CALPHAD assessments is only one of many that can be constructed from the published literature. For Nd--Fe--B, three Fe--Nd, seven Fe--B and four Nd--B assessments are freely available, giving 84 candidate ternary descriptions. We screen all 84 for feasibility of Nd\(_2\)Fe\(_{14}\)B at the stoichiometric composition over 800--1600~K and report the distribution of feasibility thresholds. Seventy-eight of the 84 combinations (92.9\,\%) are mutually compatible under a strict ($1$~\si{\joule\per\mole}) merge-then-edge-check protocol. The six failures all involve the Witusiewicz 2019 Fe--B assessment---the only one that treats boron as purely interstitial in the disordered bcc, fcc and hcp phases---and arise because the merge silently alters the sublattice model of those phases; each fails by only $6.2$~\si{\joule\per\mole} at a single test point. Re-admitting them leaves the range unchanged and shifts no median by more than $0.07$~kJ. At 800~K the ensemble threshold spans $-9.26$ to $-8.33$~\si{kJ\per\mole} (per mole of atoms; median $-8.79$, spread $0.94$). The two available full ternary assessments (Hallemans 1995 and Chen 2019) both fall inside this range, with Hallemans coinciding with the median to within $0.01$~kJ. The single-value threshold of $-9.2$~\si{kJ\per\mole} used in our prior Stage~3 work sits at the more-negative extreme of the ensemble. The spread grows from $0.94$~kJ at 800~K to $2.35$~kJ at 1600~K. We argue that the ensemble is a minimal-assumption description of the current open thermodynamic data for Nd--Fe--B, and that single-value CALPHAD claims should be reported together with the ensemble spread around them.
\end{abstract}

\section{Introduction}

The CALPHAD method has produced more than fifty years of thermodynamic assessments for binary alloy systems. For well-studied binaries, several independent assessments are typically available---each fitted against the data visible to its authors at the time, each a point estimate without a formal uncertainty. In the ternary case, where the number of binary mergers grows combinatorially with the number of available binary assessments, a user who selects a single merged description implicitly selects one point from a larger population. The spread of that population is hidden.

In earlier work on Nd--Fe--B (Radulov et al., Stage~3) we reported a feasibility threshold for Nd\(_2\)Fe\(_{14}\)B constructed from a single merged binary description based on one choice per edge. The reported value of $-9.2$~\si{kJ\per\mole} at 800~K was consistent with the two published full ternary assessments (Hallemans 1995 and Chen 2019) to within $0.1$--$0.4$~kJ at 800~K. We did not then quantify what fraction of the open-literature ensemble the single value represents, nor what the spread would be if every open combination were examined.

Here we perform that calculation. We screen all 84 binary-assessment combinations available in the SGTE binary library for the Fe--Nd--B system, apply a merge-then-edge-check protocol, and compute the Nd\(_2\)Fe\(_{14}\)B feasibility threshold for every surviving combination at eight temperatures. The resulting distribution quantifies two quantities that any single merged description leaves implicit: the mutual compatibility of the open binary data for Nd--Fe--B, and the sensitivity of the missing-ternary Nd\(_2\)Fe\(_{14}\)B feasibility threshold to the choice of binary assessment.

\section{Method}

\subsection{Assessment inventory}

The binary TDB files used in this work are part of the open SGTE collection of binary datasets~\cite{HALLSTEDT2025102833} and are also distributed, together with calculated phase diagrams, by the ORCHID project~\cite{orchid_phase_diagrams_2026,orchid_data_2026} (\url{https://data.orchid-project.org}). The SHA-256 digests of the four Nd--B files in Supplementary Table~S1a agree with those published by ORCHID. We use them as maintained in our project repository (\url{https://github.com/iliya-radulov/my_db_project}; archived release \texttt{v4.1-stage4\_trackA}: DOI \href{https://doi.org/10.5281/zenodo.24039719}{10.5281/zenodo.24039719}). For Nd--Fe--B this collection provides:
\begin{itemize}
  \item Fe--Nd: 3 assessments: \texttt{FeNd-16Che} (Chen 2016~\cite{Chen2016}), \texttt{FeNd-16Wan} (Wang 2016~\cite{Wang2016}), \texttt{FeNd-95Hal-LB} (Hallemans 1995~\cite{Hallemans1995}, as recorded in Landolt--B\"ornstein).
  \item Fe--B: 7 assessments: \texttt{FeB-01Pal} (Palumbo 2001~\cite{Palumbo2001}), \texttt{FeB-02Rom} (Van Rompaey 2002~\cite{VanRompaey2002}), \texttt{FeB-08Yos} (Yoshitomi 2008~\cite{Yoshitomi2008}), \texttt{FeB-19Wit} (Witusiewicz 2019~\cite{Witusiewicz2019}), \texttt{FeB-20Tak} (Takahashi 2020~\cite{Takahashi2020}), \texttt{FeB-91Pan} (Pan 1991, unpublished~\cite{Pan1991}), \texttt{FeB-94Hal-LB} (Hallemans 1994~\cite{Hallemans1994}, Landolt--B\"ornstein).
  \item Nd--B: 4 assessments: \texttt{NdB-16Zho-corrected} (Zhou 2016~\cite{Zhou2016}, as corrected), \texttt{NdB-19Che} (Chen 2019~\cite{Chen2019}), \texttt{NdB-95Hal-LB} (Hallemans 1995~\cite{Hallemans1995}, Landolt--B\"ornstein), \texttt{NdB-98Che} (Chevalier 1998, unpublished~\cite{Chevalier1998}).
\end{itemize}
The Cartesian product yields 84 candidate merged descriptions. The complete list of binary files and their SHA-256 digests is given in Supplementary Table~S1a so that the ensemble is exactly reproducible.

\emph{Naming convention.} Throughout, binary assessments are referred to by their file identifiers (edge, two-digit year, first letters of the first author: \texttt{FeB-19Wit}, \texttt{NdB-19Che}, \dots); author--year labels appear only in the list above. The two full ternary assessments are written \emph{Hallemans 1995 (ternary)} and \emph{Chen 2019 (ternary)} to distinguish them from the binary files of the same authors.

\subsection{Merge-then-edge-check protocol}

For each of the 84 combinations we:
\begin{enumerate}
  \item Merge the three thermodynamic database (TDB) files into one ternary description using a parsed-database merge (not text concatenation). Function and parameter conflicts are reported; shared phases are united constituent-wise.
  \item Verify that the merged description reproduces each binary file along its edge to within $1$~\si{\joule\per\mole} on the Gibbs energy of equilibrium. This tolerance is deliberately tight relative to typical CALPHAD parameter precision and to the numerical noise of the equilibrium solver, so that any detectable change in a binary edge is flagged as an inconsistency. Combinations that fail this edge check at any temperature are recorded but excluded from the subsequent feasibility screen.
  \item For the surviving combinations, compute the Nd\(_2\)Fe\(_{14}\)B feasibility threshold at eight temperatures (800, 1000, 1100, 1200, 1300, 1375, 1450, 1600~K). These span the practical processing window of Nd--Fe--B magnets (sintering, annealing and high-temperature liquid-phase regimes) while remaining computationally tractable.
\end{enumerate}
The \emph{feasibility threshold} $\Delta G_f^{*}(T)$ is the Gibbs energy of formation, relative to the SER (stable element reference) state, that a stoichiometric line compound at the Nd\(_2\)Fe\(_{14}\)B composition ($x_{\mathrm{Fe}}=14/17$, $x_{\mathrm{Nd}}=2/17$, $x_{\mathrm{B}}=1/17$) must reach to become competitive with the equilibrium phase assemblage of the merged binary description:
\begin{equation}
\Delta G_f^{*}(T) = \frac{1}{1000}\Bigl(G_{\mathrm{competing}}(T) - \sum_i x_i\,G^{\mathrm{SER}}_i(T)\Bigr)
\label{eq:threshold}
\end{equation}
in \si{kJ\per\mole} (per mole of atoms; this unit is used throughout), where $G_{\mathrm{competing}}$ is the equilibrium Gibbs energy per mole of atoms (\si{\joule\per\mole}) of the merged binary description at that composition and $G^{\mathrm{SER}}_i(T)$ are the SGTE \texttt{GHSER} reference functions of Fe, Nd and B. The compound is favoured when its own formation Gibbs energy satisfies $\Delta G_f \le \Delta G_f^{*}$. A more negative threshold therefore means that the binary phases are already strongly stable and the compound must be strongly stabilised (by ternary interactions or by its own Gibbs-energy function) to appear; a less negative threshold means that weaker stabilisation suffices. The quantity is a Gibbs-energy difference evaluated at temperature $T$; it was denoted $\Delta H_f^{*}$ in Stage~3.

The screen uses the open-source Python CALPHAD library pycalphad~\cite{Otis2017} (version 0.11.2) and our own \texttt{tdb\_merge.py} utility. The threshold kernel employs the GHSER-based reference identified as the correct one in Stage~3 \S6.2.

\subsection{Reproducibility}

Code is available under \texttt{stage4\_step1/run\_batch\_v2.py} (dependency \texttt{stage\_three/tdb\_merge.py} with the \texttt{tag} argument). Results are written per combination to \texttt{results\_v2/batch\_results.csv} and \texttt{results\_v2/detail/<combo\_id>.json}. A summary is written to \texttt{results\_v2/summary.json}. The 84 combinations were executed with 8 parallel workers. The per-combination run times sum to 7.1~min (5.1~s on average, 7.9~s at most), and the first and last combinations started $51$~s apart, so the full screen took about one minute of wall-clock time on the hardware used.

\section{Results}

\subsection{Survival}
\label{sec:survival}

Of the 84 candidate combinations, 78 (92.9\,\%) survive the merge-then-edge-check protocol. The six failures are all combinations of FeB-19Wit with NdB-19Che or NdB-16Zho-corrected (three Fe--Nd choices $\times$ two Nd--B choices). The failure mechanism is analysed in \S\ref{sec:failure}. Re-admitting the six at a relaxed tolerance (their worst deviation is $6.2$~\si{\joule\per\mole}) leaves the minimum and maximum of the threshold distribution unchanged at all eight temperatures and shifts the median by at most $0.07$~kJ (for example $-8.76 \rightarrow -8.69$ at 1600~K); the six lie between the 13th and 97th percentile of the strict ensemble. The conclusions therefore do not depend on the edge-check tolerance.

\subsection{Distribution of the feasibility threshold}

Table~\ref{tab:dist} summarises the distribution of the Nd\(_2\)Fe\(_{14}\)B feasibility threshold $\Delta G_f^{*}$ over the 78 surviving combinations at the eight temperatures.

\begin{table}[htbp]
\centering
\caption{Distribution of the feasibility threshold $\Delta G_f^{*}$ (\si{kJ\per\mole}, per mole of atoms) over the 78 survivors; spread $=\max-\min$.}
\label{tab:dist}
\begin{tabular}{rrrrrrr}
\toprule
$T$ (K) & min & q25 & median & q75 & max & spread \\
\midrule
800  & $-9.26$ & $-8.87$ & $-8.79$ & $-8.70$ & $-8.33$ & $0.94$ \\
1000 & $-8.28$ & $-8.03$ & $-7.88$ & $-7.76$ & $-7.47$ & $0.81$ \\
1100 & $-8.00$ & $-7.67$ & $-7.49$ & $-7.27$ & $-7.04$ & $0.96$ \\
1200 & $-7.81$ & $-7.36$ & $-7.17$ & $-6.76$ & $-6.58$ & $1.22$ \\
1300 & $-7.77$ & $-7.17$ & $-6.94$ & $-6.47$ & $-6.07$ & $1.70$ \\
1375 & $-8.02$ & $-7.30$ & $-7.06$ & $-6.53$ & $-5.99$ & $2.03$ \\
1450 & $-8.55$ & $-7.80$ & $-7.45$ & $-7.00$ & $-6.33$ & $2.22$ \\
1600 & $-9.87$ & $-9.34$ & $-8.76$ & $-8.38$ & $-7.51$ & $2.35$ \\
\bottomrule
\end{tabular}
\end{table}

Quartiles use linear interpolation between the 78 ordered values; per-combination values at all eight temperatures are provided in the supplementary data files.

\begin{figure}[tbp]
\centering
\includegraphics[width=0.8\textwidth]{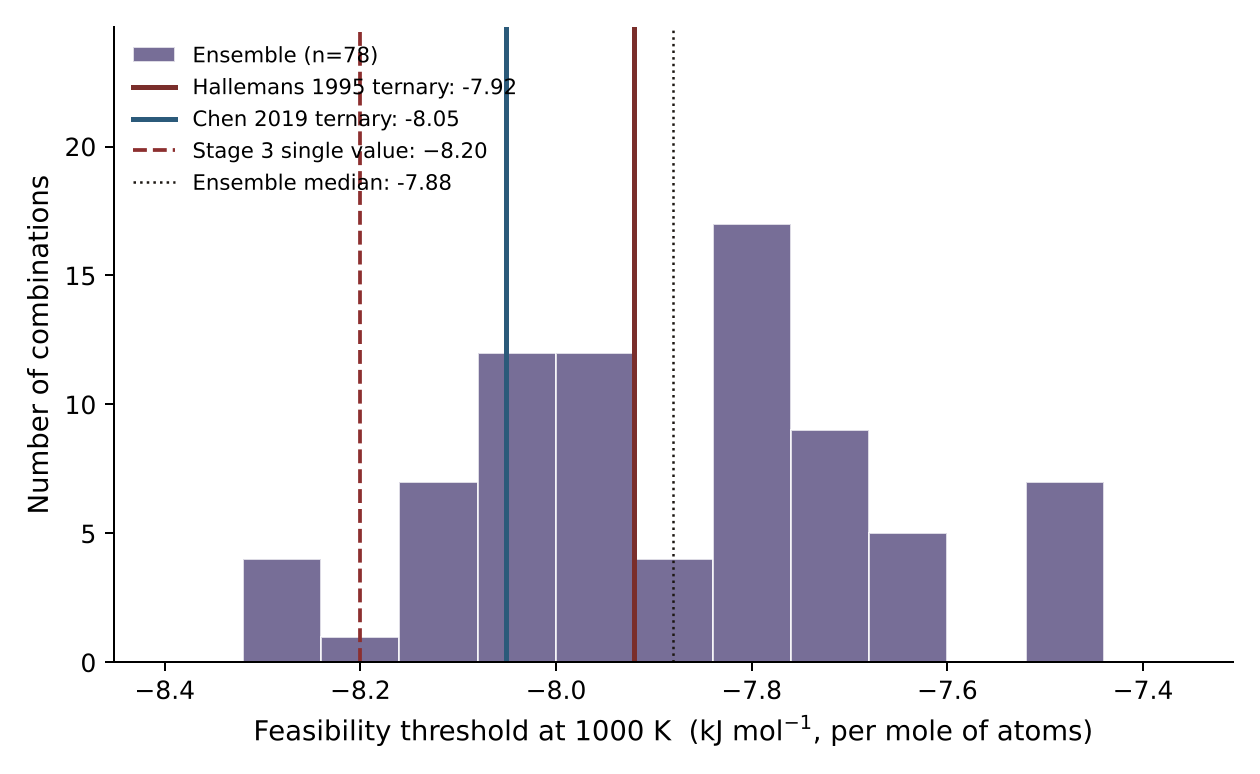}
\caption{Histogram of $\Delta G_f^{*}$ at 1000~K over the 78 survivors. Vertical lines mark the two available full ternary assessments (Hallemans 1995 and Chen 2019); the Stage~3 single value ($-8.20$ at 1000~K) is shown as a dashed line for comparison.}
\label{fig:hist}
\end{figure}

\begin{figure}[tbp]
\centering
\includegraphics[width=0.8\textwidth]{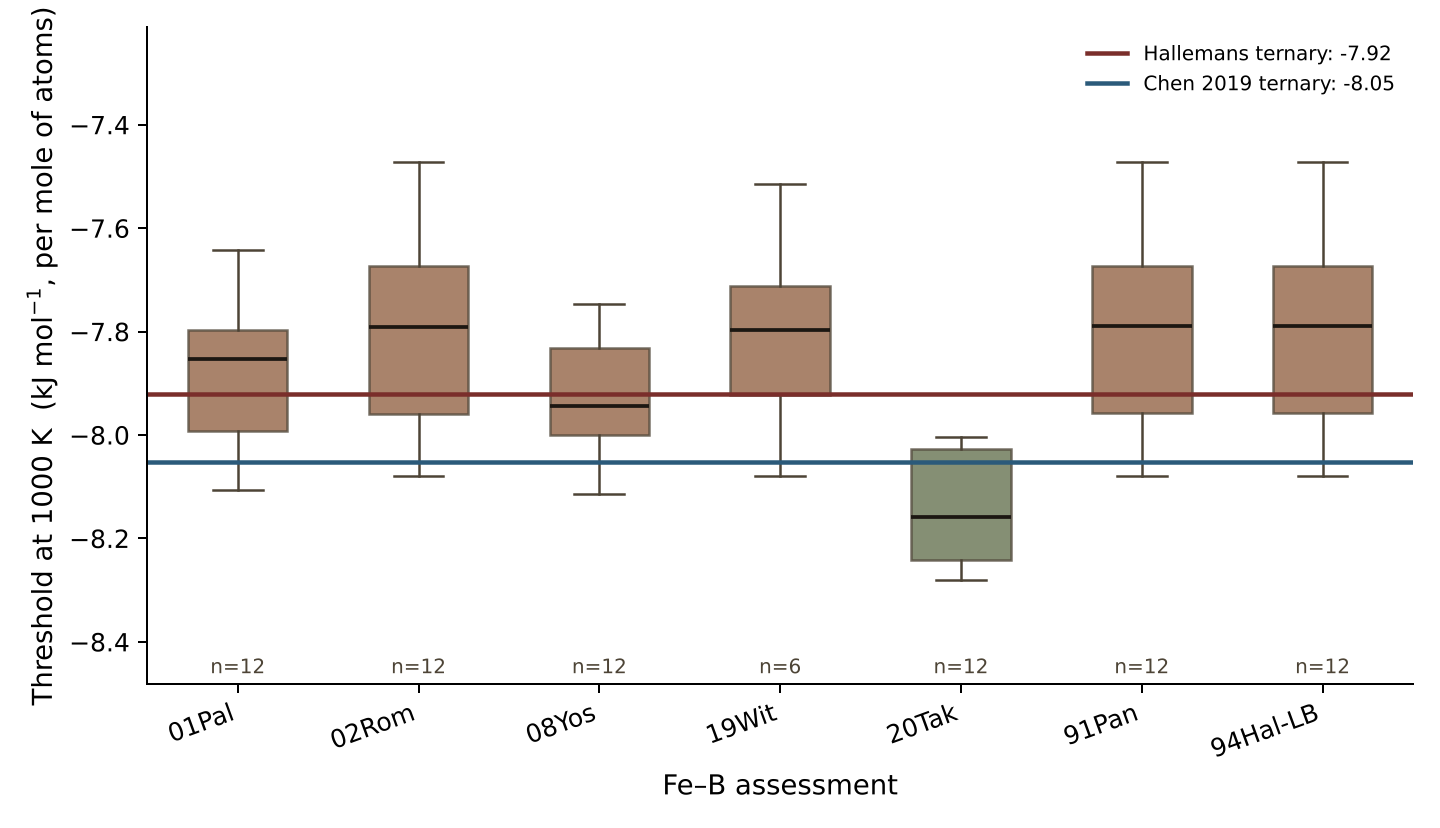}
\caption{Box plot of the ensemble grouped by Fe--B assessment at 1000~K. Takahashi 2020 produces the tightest cluster (spread $0.28$~kJ) and the most negative median ($-8.16$); the other six Fe--B assessments cluster near $-7.80$ with spreads of $0.37$--$0.61$. Takahashi's median is more negative than either ternary reference at this temperature, by $0.11$~kJ relative to Chen 2019 ($-8.05$) and $0.24$~kJ relative to Hallemans 1995 ($-7.92$). The ranking illustrates which binary edge carries the largest systematic shift on the ternary feasibility threshold.}
\label{fig:box_feb}
\end{figure}

The spread grows by a factor of approximately 2.5 between 800~K and 1600~K, reflecting the loss of calorimetric constraint at high temperatures and the accumulated extrapolation uncertainty of the binary assessments. To see which edge drives the spread we fitted, at each temperature, an additive model with the three edge choices as categorical factors and report each factor's sum of squares as a fraction of the total variance; the remainder is the interaction term (4\,\% at 800~K, 7\,\% at 1000~K, at most 1\,\% from 1300~K). The Fe--Nd choice explains the largest share at 800~K (51\,\%, against 26\,\% for Fe--B and 20\,\% for Nd--B). At 1000~K, the temperature of Figs.~\ref{fig:hist} and~\ref{fig:box_feb}, the shares are 30, 36 and 26\,\% (Fe--Nd, Fe--B, Nd--B), and the Fe--B choice dominates from 1300~K upward (67, 58, 56 and 66\,\% at 1300, 1375, 1450 and 1600~K).

\subsection{Comparison with full ternary assessments}

We applied the same feasibility screen to the two full ternary CALPHAD assessments available in the open literature for Nd--Fe--B: Hallemans 1995 (ternary~\cite{Hallemans1995}; \texttt{bfend\_hal\_en.tdb}) and Chen 2019 (ternary~\cite{Chen2019}; \texttt{Nd-Fe-B.TDB}). We also screened the NIMS Nd-mag-STAM multicomponent database as a cross-check; its reported Nd\(_2\)Fe\(_{14}\)B threshold is identical to Hallemans 1995 to $10^{-4}$~kJ at every temperature. This confirms that the STAM database inherits its Fe--Nd--B subsystem from Hallemans 1995 and does not constitute a third independent ternary assessment. We therefore report Hallemans 1995 and Chen 2019 as the two independent references. For both assessments the native Nd\(_2\)Fe\(_{14}\)B phase was suppressed so that the computed threshold is directly comparable to the binary-ensemble threshold.

\begin{table}[htbp]
\centering
\caption{Full ternary assessments placed on the ensemble distribution (\si{kJ\per\mole}, per mole of atoms).}
\label{tab:ternary}
\begin{tabular}{rrrrr}
\toprule
$T$ (K) & Hallemans 1995 & Chen 2019 & Ensemble median & Ensemble range \\
\midrule
800  & $-8.78$ & $-9.10$ & $-8.79$ & $-9.26 \ldots -8.33$ \\
1000 & $-7.92$ & $-8.05$ & $-7.88$ & $-8.28 \ldots -7.47$ \\
1100 & $-7.51$ & $-7.53$ & $-7.49$ & $-8.00 \ldots -7.04$ \\
1200 & $-7.12$ & $-7.04$ & $-7.17$ & $-7.81 \ldots -6.58$ \\
1300 & $-6.75$ & $-6.56$ & $-6.94$ & $-7.77 \ldots -6.07$ \\
1375 & $-6.73$ & $-6.41$ & $-7.06$ & $-8.02 \ldots -5.99$ \\
1450 & $-7.20$ & $-6.75$ & $-7.45$ & $-8.55 \ldots -6.33$ \\
1600 & $-8.57$ & $-8.01$ & $-8.76$ & $-9.87 \ldots -7.51$ \\
\bottomrule
\end{tabular}
\end{table}

Three features of the comparison are worth emphasising.
\begin{enumerate}
  \item \textbf{At low temperature, both ternary assessments sit inside the ensemble bulk.} At 800~K and 1000~K the Hallemans 1995 ternary coincides with the ensemble median to within $0.01$ and $0.04$~\si{kJ\per\mole} respectively. The Chen 2019 ternary sits approximately one histogram-bin to the more-negative side of the median ($-9.10$ versus $-8.79$ at 800~K). Both references fall well inside the ensemble range. The ensemble is therefore internally calibrated against an independent full ternary assessment; the median is not merely a statistical convenience but coincides quantitatively with what a careful ternary fit of the same open data produces.
  \item \textbf{The Stage~3 single-point value sits at the more-negative extreme of the ensemble.} The Stage~3 paper reported $-9.20$~\si{kJ\per\mole} at 800~K from a single merged binary description. In the present ensemble this value lies in the most-negative histogram bin of the distribution---between the Chen 2019 ternary reference ($-9.10$) and the ensemble minimum ($-9.26$)---and $0.41$~kJ below the ensemble median ($-8.79$). A typical open-assessment combination therefore yields a threshold approximately $0.4$~kJ less negative than the value previously reported. The Stage~3 result was an honest consequence of the specific binary choice made at the time, but it was not a typical draw from the open literature. Future readers of the Stage~3 paper should carry this context.
  \item \textbf{At high temperature the ensemble is more pessimistic than both ternaries.} From 1200~K upward both ternary thresholds lie above (are less negative than) the ensemble median at every temperature computed: Hallemans 1995 by $0.05$--$0.33$~kJ and Chen 2019 by $0.13$--$0.75$~kJ between 1200 and 1600~K (Table~\ref{tab:ternary}; for example $-6.73$ versus $-7.06$ at 1375~K and $-8.57$ versus $-8.76$ at 1600~K). At 800--1100~K, in contrast, Hallemans lies within $0.04$~kJ of the median and Chen at or below it, so the crossover lies between 1100 and 1200~K, where both ternaries sit within $0.13$~kJ of the median. The offset grows with temperature and is largest for Chen 2019, which lies at the 81st--87th percentile of the ensemble across 1375--1600~K, reaching the 87th at 1600~K.

  Since a less negative threshold means weaker stabilisation suffices, the binary ensemble is more pessimistic about Nd\(_2\)Fe\(_{14}\)B stability at high temperature than the full ternary assessments. We interpret this as a modelling asymmetry rather than a disagreement to be reconciled: the full ternary assessments carry a Nd\(_2\)Fe\(_{14}\)B-specific Gibbs-energy function that smooths the high-temperature behaviour of the compound explicitly, whereas the binary ensemble treats Nd\(_2\)Fe\(_{14}\)B as a hypothetical line compound and is therefore more sensitive to the high-temperature extrapolation of the binary assessments. Both framings are internally consistent. The ensemble's picture of uncertainty is itself temperature-dependent---its spread grows by a factor of 2.5 between 800~K and 1600~K---and it appropriately flags a single-value claim at high temperature as carrying more hidden uncertainty than one at low temperature.
\end{enumerate}

\begin{figure}[tbp]
\centering
\includegraphics[width=0.8\textwidth]{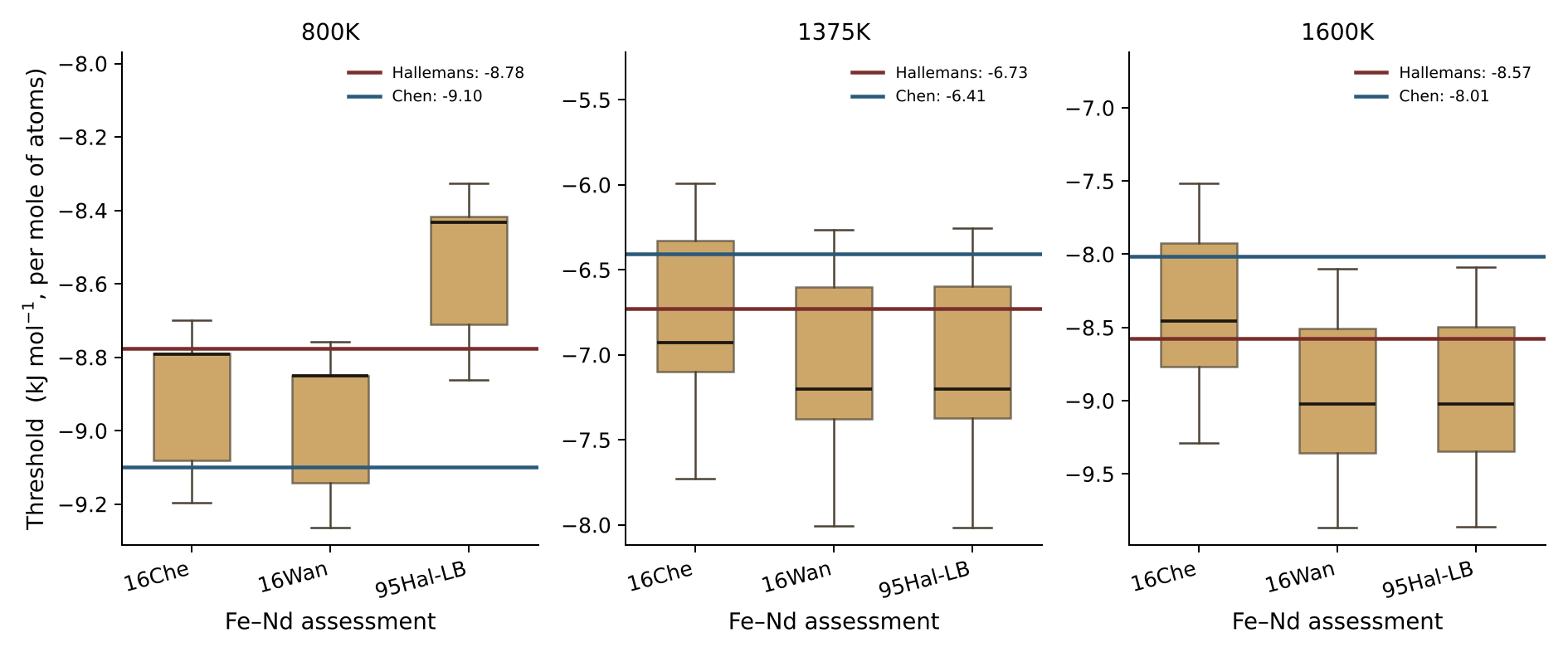}
\caption{Box plot of the ensemble grouped by Fe--Nd assessment at three temperatures (800, 1375 and 1600~K). Hallemans 1995 (ternary) and Chen 2019 (ternary) references are overlaid as horizontal lines.}
\label{fig:box_fend}
\end{figure}

\subsection{Failure mechanism}
\label{sec:failure}

\begin{figure}[tbp]
\centering
\includegraphics[width=\textwidth]{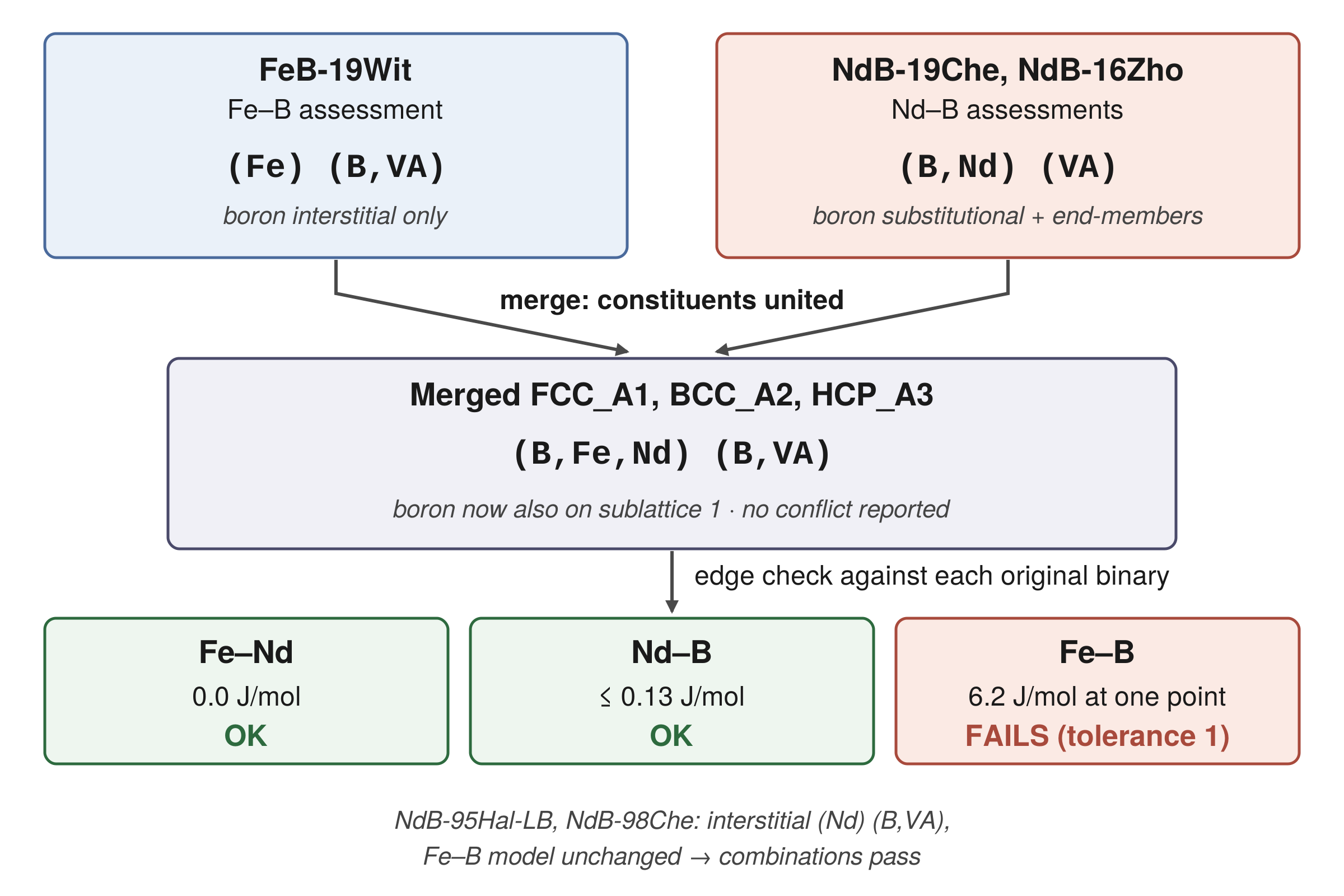}
\caption{Schematic of the failure mechanism. \texttt{FeB-19Wit} models boron in the bcc, fcc and hcp phases as purely interstitial, $(\mathrm{Fe})(\mathrm{B},\mathrm{VA})$; \texttt{NdB-19Che} and \texttt{NdB-16Zho-corrected} model it as substitutional, $(\mathrm{B},\mathrm{Nd})(\mathrm{VA})$. The merge unites the constituents, giving $(\mathrm{B},\mathrm{Fe},\mathrm{Nd})(\mathrm{B},\mathrm{VA})$ together with the Nd--B end-member parameters, without reporting a conflict. The Fe--Nd and Nd--B edges are reproduced, but the Fe--B edge deviates by $6.2$~\si{\joule\per\mole} at one test point, exceeding the $1$~\si{\joule\per\mole} tolerance. The two Nd--B files with an interstitial model (\texttt{NdB-95Hal-LB}, \texttt{NdB-98Che}) leave the Fe--B model unchanged. The additional \texttt{MB\_B33} polymorph of \texttt{FeB-19Wit} does not contribute (text).}
\label{fig:mechanisms}
\end{figure}

The six failures are all combinations of \texttt{FeB-19Wit} (Witusiewicz 2019) with either \texttt{NdB-19Che} (Chen 2019) or \texttt{NdB-16Zho-corrected} (Zhou 2016, corrected). \texttt{FeB-19Wit} differs from the other Fe--B assessments in two ways; only the second causes failures (Fig.~\ref{fig:mechanisms}). The full diagnostic chain, including the two hypotheses that were tested and rejected, is documented in Supplementary Information~S2.

\textbf{Structural difference --- an additional Fe--B polymorph.} \texttt{FeB-19Wit} models the Fe--B mono-boride as two crystallographic polymorphs, \texttt{MB\_B27} (FeB-type, \textit{Pnma}) and \texttt{MB\_B33} (CrB-type, \textit{Cmcm}); all six other Fe--B assessments model only \texttt{MB\_B27}. The extra phase is carried into every merged description containing \texttt{FeB-19Wit}, but it does not by itself cause an edge-check failure: all six combinations of \texttt{FeB-19Wit} with \texttt{NdB-95Hal-LB} or \texttt{NdB-98Che} pass, and the merged descriptions of those six reproduce the \texttt{FeB-19Wit} edge within tolerance. The extra polymorph therefore makes the \texttt{FeB-19Wit} ensemble members structurally different, but it is not the source of the failures.

\textbf{Failure mechanism --- silent change of the phase model.} The binary assessments use two incompatible sublattice models for boron in the shared phases FCC\_A1, BCC\_A2 and HCP\_A3 (Table~\ref{tab:mech2}): a \emph{substitutional} model, $(\mathrm{B},M)(\mathrm{VA})$ with $M=$ Fe or Nd, and an \emph{interstitial} model, $(M)(\mathrm{B},\mathrm{VA})$. \texttt{FeB-19Wit} is the only Fe--B file using the interstitial form; the other six use the substitutional form. Among the Nd--B files, \texttt{NdB-19Che} and \texttt{NdB-16Zho-corrected} use the substitutional form and \texttt{NdB-95Hal-LB} and \texttt{NdB-98Che} the interstitial form. The parsed-database merge unites the constituents of shared phases. Merging \texttt{FeB-19Wit} with a substitutional Nd--B file therefore turns, for example, FCC\_A1 from $(\mathrm{Fe})(\mathrm{B},\mathrm{VA})$ into $(\mathrm{B},\mathrm{Fe},\mathrm{Nd})(\mathrm{B},\mathrm{VA})$: boron gains a position on the first sublattice, together with the end-member parameters supplied by the Nd--B file (e.g.\ \texttt{G(FCC\_A1,B:VA;0)} $=43514+\mathrm{GHSERBB}-12.217\,T$~\si{\joule\per\mole}), which \texttt{FeB-19Wit} alone does not contain. The merge adds four such parameters on the Fe--B edge, redefines no symbol of \texttt{FeB-19Wit}, and reports no conflict for them. The Fe--Nd and Nd--B edges are reproduced to $\le 0.13$~\si{\joule\per\mole}, but the Fe--B edge deviates by $6.21$~\si{\joule\per\mole} at one of 30 test points (1600~K, $x_{\mathrm{Fe}}=0.95$, FCC\_A1\,+\,liquid in both descriptions), with no phase-set mismatch; all six failing combinations show the same deviation at the same point. Restoring the constituents of \texttt{FeB-19Wit} on the three merged phases removes the deviation completely ($0.0$~\si{\joule\per\mole} at all 30 points, with or without the four added parameters), whereas deleting only the four added parameters leaves $2.5$~\si{\joule\per\mole}. The unified constituent set is therefore the cause, and the added end-member parameters are part of its effect. The change is silent because the merge reports it neither as a conflict nor as an error; it is visible only to an edge check against the original binary.

The same union produces a mixed model in 36 of the 78 survivors (a substitutional Fe--B file with an interstitial Nd--B file); these pass the edge check, so their induced energy change is below the tolerance, but they are not validated beyond it.

\begin{table}[htbp]
\centering
\caption{Boron sublattice model of the four Nd--B files and outcome of the merge with \texttt{FeB-19Wit}, whose own model is interstitial, $(\mathrm{Fe})(\mathrm{B},\mathrm{VA})$.}
\label{tab:mech2}
\begin{tabular}{lll}
\toprule
Nd--B file & Boron model in FCC\_A1, BCC\_A2, HCP\_A3 & Outcome with \texttt{FeB-19Wit} \\
\midrule
NdB-19Che          & substitutional $(\mathrm{B},\mathrm{Nd})(\mathrm{VA})$ & fails \\
NdB-16Zho-corrected & substitutional $(\mathrm{B},\mathrm{Nd})(\mathrm{VA})$ & fails \\
NdB-95Hal-LB       & interstitial $(\mathrm{Nd})(\mathrm{B},\mathrm{VA})$ & succeeds \\
NdB-98Che          & interstitial $(\mathrm{Nd})(\mathrm{B},\mathrm{VA})$ & succeeds \\
\bottomrule
\end{tabular}
\end{table}

Neither feature is a bug in pycalphad or in the published assessments considered individually. Each assessment is internally self-consistent. The failure arises from combining two assessments that model boron on different sublattices, and it would be invisible to any single-assessment user; it is surfaced only by the ensemble protocol.

\section{Discussion}

\subsection{The Stage~3 single value, reinterpreted}

In Stage~3 we reported a feasibility threshold of $-9.2$~\si{kJ\per\mole} at 800~K from a single merged binary description. The present ensemble places that value at the more-negative extreme of the 78-member distribution, just inside the range ($-9.26$ to $-8.33$) and $0.4$~kJ below the median ($-8.79$). The Stage~3 value was therefore an honest result of the specific binary choice made at the time, but it was not a typical draw from the open literature. A typical open-assessment combination gives a threshold about $0.4$~kJ less negative---closer to the Hallemans 1995 (ternary) than to the Chen 2019 (ternary). Future readers of the Stage~3 paper should carry this context.

\subsection{Hallemans at the median}

The quantitative result most useful for calibration is the coincidence of Hallemans 1995 with the ensemble median. At 800~K the two agree to within $0.01$~kJ. We do not claim that this reflects a particular virtue of Hallemans's assessment; the ensemble median is simply what the open binary data say on average, and a full ternary assessment that fits the same data should agree with it. The coincidence is strong evidence that the ensemble is internally calibrated against an independent full ternary assessment, not merely a parameterisation exercise.

\subsection{High-temperature pessimism}

From 1200~K upward both ternary assessments lie above (are less negative than) the ensemble median, and at 1375 to 1600~K Chen 2019 (ternary) sits at the 81st--87th percentile of the ensemble, reaching the 87th at 1600~K. Because a less negative threshold means weaker stabilisation suffices, the ensemble is more pessimistic about Nd\(_2\)Fe\(_{14}\)B stability at high temperature. This is not a disagreement to be reconciled---both framings are internally consistent---but it is a useful reminder that the ensemble's picture of uncertainty is itself temperature-dependent. The spread grows by a factor of 2.5 between 800~K and 1600~K; a single-value claim at high temperature carries more hidden uncertainty than a single-value claim at low temperature.

\subsection{Failures as a result}

The six failures are a result, not an artefact, although each is a small energy mismatch (for the combination examined in detail, $6$~\si{\joule\per\mole} at a single test point) rather than a gross inconsistency. They identify \texttt{FeB-19Wit} as the only Fe--B assessment that models boron as purely interstitial in the bcc, fcc and hcp phases, and \texttt{NdB-19Che} and \texttt{NdB-16Zho-corrected} as the Nd--B members that model it as substitutional. The ensemble method surfaces these authoring inconsistencies between published open assessments; a single-user workflow would never see them. This is a modest but useful methodological byproduct of the approach.

\subsection{Limitations}

\begin{itemize}
  \item The ensemble is binary-only. Full ternary interactions are not modelled.
  \item The composition is pinned at exact Nd\(_2\)Fe\(_{14}\)B. The screen is a line-compound stability check against the merged binary assemblage; it is not a full ternary phase-diagram calculation.
  \item The merge unites the constituents of shared phases, which silently changes the sublattice model whenever the two files disagree (36 survivors combine a substitutional Fe--B with an interstitial Nd--B file); the edge check bounds the effect on the binary edges but not on the ternary threshold.
  \item The $1$~\si{\joule\per\mole} edge-check tolerance is strict; the six excluded combinations deviate by at most $6.2$~\si{\joule\per\mole}. Re-admitting them does not change the conclusions (\S\ref{sec:survival}), but the headline survival rate of 92.9\,\% depends on this tolerance.
  \item The 78 survivors are treated as an unweighted population. A weighting by data quality, year, or by a posterior statistical treatment (as outlined in our ongoing methodological work) would sharpen the distribution.
  \item The library of open Nd--Fe--B assessments is modest ($3 \times 7 \times 4 = 84$ combinations, 78 surviving). For other alloy systems the available combinations may be larger or smaller.
\end{itemize}

\section{Code and data availability}

The batch script, merge utility and threshold kernel are available at \url{github.com/iliya-radulov/my_db_project} under \texttt{stage\_three/stage4\_step1/}. The SGTE binary assessment library is referenced in Supplementary Table~S1a, and the ternary files used for the overlay in Table~S1b, with SHA-256 digests. The diagnostic scripts of Supplementary Information~S2 are in the subfolder \texttt{diagnostics/}. The per-combination output files (CSV, JSON details and merged TDBs) are included in the archived release (release \texttt{v4.1-stage4\_trackA}, DOI \href{https://doi.org/10.5281/zenodo.24039719}{10.5281/zenodo.24039719}; all versions of the repository: \href{https://doi.org/10.5281/zenodo.23155920}{10.5281/zenodo.23155920}).

\noindent\textbf{Implementation note.} A shared intermediate-filename race condition in \texttt{tdb\_merge.py} under parallel execution was identified and fixed by introducing a \texttt{tag} argument, so that each of the 84 combinations writes to its own named file. Generally, parallel workers must never share a fixed intermediate filename.

\section{Conclusions}

Treating the open binary CALPHAD assessments of Nd--Fe--B as a population rather than selecting one combination yields a feasibility-threshold distribution for Nd\(_2\)Fe\(_{14}\)B that is $0.94$~kJ wide at 800~K and $2.35$~kJ wide at 1600~K, and that contains both available full ternary assessments, with Hallemans 1995 (ternary) at the median at low temperature. The single-value threshold used in our prior work sits at the more-negative extreme of this distribution. The six-combination failure rate (7\,\%) is a tolerance-level result: it arises only when the single Fe--B assessment with an interstitial boron model is merged with the two Nd--B assessments with a substitutional one, and including the six combinations changes the distribution negligibly. The ensemble is a minimal-assumption description of the current open thermodynamic data for Nd--Fe--B.

\section*{Acknowledgements}

We would like to thank Hongbin Zhang for fruitful discussions. We also acknowledge the use of Claude Sonnet 5.5 (Anthropic) in the preparation of this manuscript and of the accompanying code; the author reviewed the final content and retains full responsibility for it.

\bibliographystyle{unsrt}
\bibliography{refs}

\end{document}


\section*{An ensemble of open binary CALPHAD assessments quantifies the uncertainty of the $\mathrm{Nd}_{2}\mathrm{Fe}_{14}\mathrm{B}$ feasibility threshold and surfaces inconsistencies invisible to single-assessment users}

\medskip

\subsection*{Supplementary Information S1}

\medskip

\subsection*{ Binary CALPHAD assessments used in the ensemble and their SHA-256 digests.}

\medskip

\noindent\textbf{Table S1a. Binary assessments}

\renewcommand{\arraystretch}{1.6} 
\begin{longtable}{@{}p{1.2cm} p{3.9cm} p{3.4cm} p{5.6cm}@{}}
\toprule
Edge & File ID (\texttt{.tdb}) & Source & SHA-256 \\
\midrule
\endfirsthead
\toprule
Edge & File ID (\texttt{.tdb}) & Source & SHA-256 \\
\midrule
\endhead
\bottomrule
\endfoot

Fe--Nd & \texttt{FeNd-16Che} & Chen 2016 & {\scriptsize\texttt{\seqsplit{e50d3fee1b48c398f2a24f8c5a181a40b45eb4a8acc42b184e1824fdda0fe6ff}}} \\
Fe--Nd & \texttt{FeNd-16Wan} & Wang 2016 & {\scriptsize\texttt{\seqsplit{fa02ac1db21fdc42395a2ebd7918221fea9bf7fd924ce6942972e977681649e4}}} \\
Fe--Nd & \texttt{FeNd-95Hal-LB} & Hallemans 1995 (Landolt--B\"ornstein) & {\scriptsize\texttt{\seqsplit{ab4a9e90b16a414a4eb6d9d39cae634150efd7762fdd236541a9e2b623a8f870}}} \\
Fe--B & \texttt{FeB-01Pal} & Palumbo 2001 & {\scriptsize\texttt{\seqsplit{d868b8575842d60b902a516560a74f20e76d051efc995445aa9be40f1d01aabf}}} \\
Fe--B & \texttt{FeB-02Rom} & Van Rompaey 2002 & {\scriptsize\texttt{\seqsplit{76003da4f9d7a948978d07416cd8c6fc6849efd95bd33ef34a5216ede742c992}}} \\
Fe--B & \texttt{FeB-08Yos} & Yoshitomi 2008 & {\scriptsize\texttt{\seqsplit{bca7bc67b25d3e8bfbee09d03e47ecdb901b39084cb66856cc2a725a070ee6f1}}} \\
Fe--B & \texttt{FeB-19Wit} & Witusiewicz 2019 & {\scriptsize\texttt{\seqsplit{a20e39e4c8231b9ef69fed71aa7c23195d88d62d6db3aacc248fcd404445df70}}} \\
Fe--B & \texttt{FeB-20Tak} & Takahashi 2020 & {\scriptsize\texttt{\seqsplit{17fba6129d08f63e0d3725813578443a16b5ed00be2a1edca5cad04e5fe9315c}}} \\
Fe--B & \texttt{FeB-91Pan} & Pan 1991 & {\scriptsize\texttt{\seqsplit{9f52784aa22c8ff3f0897df2ad6878b58c903241f0983e5430d5ca3db77f2e44}}} \\
Fe--B & \texttt{FeB-94Hal-LB} & Hallemans 1994 (Landolt--B\"ornstein) & {\scriptsize\texttt{\seqsplit{724a628400ef810cb479abfadf4cb9931f7b146840942f0561e61a99cf63763b}}} \\
Nd--B & \texttt{NdB-16Zho-corrected} & Zhou 2016 (corrected) & {\scriptsize\texttt{\seqsplit{cae81a496f38b11baf672c692c03e03d65f243a869a4f0715776789e813528e3}}} \\
Nd--B & \texttt{NdB-19Che} & Chen 2019 & {\scriptsize\texttt{\seqsplit{dff8ac7c4ad64e6cd77451678648ddc632f81955b472bdbe3a41284d537c98af}}} \\
Nd--B & \texttt{NdB-95Hal-LB} & Hallemans 1995 (Landolt--B\"ornstein) & {\scriptsize\texttt{\seqsplit{a14f51e14303dba1076678e971c46f3a5a25cd24659c1aaedf8f6e472e183a1b}}} \\
Nd--B & \texttt{NdB-98Che} & Chevalier 1998 (unpublished) & {\scriptsize\texttt{\seqsplit{502e0edcb7d117b4a2497ba7d44c99b5143944f45175f4c54963b4d25486498a}}} \\
\end{longtable}

\newpage

\noindent\textbf{Table S1b. Full ternary assessments and multicomponent database used for the overlay}

\renewcommand{\arraystretch}{1.6}
\begin{longtable}{@{}p{3.0cm} p{3.4cm} p{3.4cm} p{4.4cm}@{}}
\toprule
Role & File (\texttt{.tdb}) & Source & SHA-256 \\
\midrule
\endfirsthead
\toprule
Role & File (\texttt{.tdb}) & Source & SHA-256 \\
\midrule
\endhead
\bottomrule
\endfoot

Hallemans 1995 (ternary) & \texttt{bfend\_hal\_en.tdb} & OpenCALPHAD v6 examples, B--Fe--Nd & {\scriptsize\texttt{\seqsplit{33b120c533df5d28b09f193d9d0b0eedc1970dc9fdae76202eda9fa87c802f1f}}} \\
Chen 2019 (ternary) & \texttt{Nd-Fe-B.TDB} & pycalphad v0.10.3 examples, B--Fe--Nd & {\scriptsize\texttt{\seqsplit{a9fcd84995e0b7bae9b7cbaf7b9b27e05cce07d731d5c6f018903d19d02ac4ac}}} \\
Cross-check only (NIMS Nd-mag-STAM) & \texttt{Nd-mag-STAM\_en.TDB} & OpenCALPHAD v6 examples, B--Fe--Nd & {\scriptsize\texttt{\seqsplit{2da8028187d8553d3106ca455376f253cc337278c47dbcbda60cdf0240af083c}}} \\
\end{longtable}

\noindent\emph{Note.} A second copy of the Chen 2019 file (\texttt{old\_tdb/Nd-Fe-B.TDB}, SHA-256 {\scriptsize\texttt{\seqsplit{97074741db5bce569544346c92548a689dbf0b0b090f95adac492404cdbad89c}}}) has a different hash but gives identical thresholds at all eight temperatures (800~K: $-9.0981$~kJ\,mol$^{-1}$; 1600~K: $-8.0143$~kJ\,mol$^{-1}$). The thresholds of Hallemans 1995 (ternary) and of the Nd-mag-STAM database agree to $10^{-4}$~kJ\,mol$^{-1}$ at every temperature (800~K: $-8.7757$).

\subsection*{Nd--B}

\begin{itemize}
\item Nd--B [342] P.-Y. Chevalier Unpublished Work (1998)
\item Nd--B [1007] B. Hallemans, P. Wollants, J.R. Roos, Thermodynamic assessment of the Fe-Nd-B phase diagram, \emph{J. Phase Equilib}, 16 (1995), pp. 137-149, \url{https://doi.org/10.1007/BF02664851}
\item Nd--B [1492] T.L. Chen, J. Wang, C. Guo, C. Li, Z. Du, G. Rao, H.Y. Zhou, Thermodynamic description of the Nd-Fe-B ternary system, \emph{Calphad}, 66 (2019), p. 101627, \url{https://doi.org/10.1016/j.calphad.2019.101627}
\item Nd--B [1493] G. Zhou, Y. Luo, Y. Zhou, Thermodynamic reassessment of the Nd-Fe-B ternary system, \emph{J. Electron. Mater}, 45 (2016), pp. 418-425, \url{https://doi.org/10.1007/s11664-015-4112-y}
\end{itemize}

\subsection*{Fe--B}

\begin{itemize}
\item Fe--B [711] L.-M. Pan Unpublished Work (1991)
\item Fe--B [954] K. Takahashi, K. Ishikawa, M. Fujioka, M. Enoki, H. Ohtani, Thermodynamic analysis of the Fe-Mo-B ternary system, \emph{Tetsu-to-Hagane}, 106 (2020), pp. 310-320, \url{https://doi.org/10.2355/tetsutohagane.TETSU-2019-097}
\item Fe--B [955] V.T. Witusiewicz, A.A. Bondar, U. Hecht, A. Theofilatos, N.I. Tsyganenko, S.V. Utkin, I.B. Tikhonova, Experimental study and thermodynamic re-modelling of the constituent binaries and ternary B--Fe--Ti system, \emph{J. Alloys Compd}, 800 (2019), pp. 419-449, \url{https://doi.org/10.1016/j.jallcom.2019.05.341}
\item Fe--B [956] K. Yoshitomi, Y. Nakama, H. Ohtani, M. Hasebe, Thermodynamic analysis of the Fe--Nb--B ternary system, \emph{ISIJ Int}, 48 (2008), pp. 835-844, \url{https://doi.org/10.2355/isijinternational.48.835}
\item Fe--B [957] T. Van Rompaey, K.C. Hari Kumar, P. Wollants, Thermodynamic optimization of the B--Fe system, \emph{J. Alloys Compd.}, 334 (2002), pp. 173-181, \url{https://doi.org/10.1016/S0925-8388(01)01777-7}
\item Fe--B [958] M. Palumbo, G. Cacciamani, E. Bosco, M. Baricco, Thermodynamic analysis of glass formation in Fe-B system, \emph{Calphad}, 25 (2001), pp. 625-637, \url{https://doi.org/10.1016/S0364-5916(02)00012-3}
\item Fe--B [959] B. Hallemans, P. Wollants, J.R. Roos, Thermodynamic reassessment and calculation of the Fe-B phase diagram, \emph{Z. Metallkd}, 85 (1994), pp. 676-682, \url{https://doi.org/10.1515/ijmr-1994-851002}
\end{itemize}

\subsection*{Fe--Nd}

\begin{itemize}
\item Fe--Nd [1005] W. Wang, T. Liu, H. Song, S. Xue, W. Zhang, H. Zhang, L. Zhang, Y. Jia, N. Wang, X. Cheng, C. Zeng, A thermodynamic modeling of the Fe-Nd-Sb system, \emph{Mater. Trans.} (JIM), 57 (2016), pp. 103-111, \url{https://doi.org/10.2320/matertrans.M2015333}
\item Fe--Nd [1006] T.L. Chen, J. Wang, M.H. Rong, G. Rao, H.Y. Zhou, Experimental investigation and thermodynamic assessment of the Fe-Pr and Fe-Nd binary systems, \emph{Calphad}, 55 (2016), pp. 270-280, \url{https://doi.org/10.1016/j.calphad.2016.10.004}
\item Fe--Nd [1007] B. Hallemans, P. Wollants, J.R. Roos, Thermodynamic assessment of the Fe-Nd-B phase diagram, \emph{J. Phase Equilib}, 16 (1995), pp. 137-149, \url{https://doi.org/10.1007/BF02664851}
\end{itemize}


\section*{An ensemble of open binary CALPHAD assessments quantifies the uncertainty of the $\mathrm{Nd}_{2}\mathrm{Fe}_{14}\mathrm{B}$ feasibility threshold and surfaces inconsistencies invisible to single-assessment users}

\medskip

\subsection*{Supplementary Information S2}

\medskip

\subsection*{ Diagnostic Analysis of the $\mathrm{Nd}_{2}\mathrm{Fe}_{14}\mathrm{B}$ Feasibility Threshold}

\medskip

The six combinations that fail the $1~\mathrm{J\,mol^{-1}}$ edge check all contain \texttt{FeB-19Wit} together with either \texttt{NdB-19Che} or \texttt{NdB-16Zho-corrected}. The following sequence of diagnostic scripts was used to identify the origin of the failures. \\[0.3em]
All scripts are available in the repository at: \url{https://github.com/iliya-radulov/my_db_project}\\[0.3em]
commit \texttt{e373cdc83c39af599b8df1e8222f6bedf4cf90db} under \texttt{stage\_three/stage4\_step1/diagnostics/}.\\[0.3em]
The archived release \texttt{v4.0-stage4\_trackA} has \href{https://doi.org/10.5281/zenodo.24034300}{DOI:10.5281/zenodo.24034300}.\\[0.3em]

\renewcommand{\arraystretch}{1.6} 
\begin{longtable}{@{}p{0.5cm} p{4.2cm} p{4.2cm} p{5.2cm}@{}}
\toprule
Step & Script & Question addressed & Outcome \\
\midrule
\endfirsthead
\toprule
Step & Script & Question addressed & Outcome \\
\midrule
\endhead
\bottomrule
\endfoot

1 & \texttt{extract\_all\_T.py} & Thresholds of the 78 survivors at all eight temperatures & Written to \texttt{results\_v2/ensemble\_all\_T.csv} \\
2 & \texttt{which\_edge\_fails.py} & Which binary edge fails for the representative combination \texttt{FeNd-16Che} / \texttt{FeB-19Wit} / \texttt{NdB-19Che}? & Fe--B edge: $6.2~\mathrm{J\,mol^{-1}}$ at 1600~K, $x_{\mathrm{Fe}}=0.95$; Fe--Nd $0.0~\mathrm{J\,mol^{-1}}$; Nd--B $\le 0.1~\mathrm{J\,mol^{-1}}$ \\
3 & \texttt{sensitivity\_relaxed.py} & Do the six near-misses alter the reported distribution? & Minimum and maximum unchanged at every temperature; median shift $\le 0.07~\mathrm{kJ}$ \\
4 & \texttt{diag\_fcc\_B\_v2.py} & Do the Nd--B files reference the symbols \texttt{GBCCBB} / \texttt{GHCPBB}? & No. End-members are numeric $+$ \texttt{GHSERBB} \\
5 & \texttt{merge\_conflicts.py} & What parameter conflicts does the merge report? & Only \texttt{G(FCC\_A1,ND:VA;0)}, also present in passing combinations \\
6 & \texttt{fe\_b\_edge\_additions.py} & What does the merge add on the Fe--B edge? & Four additional parameters; no symbol is redefined \\
7 & \texttt{test\_remove\_added.py} & Are those four parameters responsible for the deviation? & No: $2.5~\mathrm{J\,mol^{-1}}$ residual remains \\
8 & \texttt{phase\_models.py} & Are differences in magnetic model constants the cause? & No: constants identical across files \\
9 & \texttt{constituent\_union\_test.py} & Is the united sublattice model of the disordered phases the cause? & Yes. Restoring the original \texttt{FeB-19Wit} constituents reduces the deviation to $0.0~\mathrm{J\,mol^{-1}}$ \\
\end{longtable}

\subsection*{Rejected hypotheses (retained for the record)}

\begin{itemize}
\item Extra \texttt{MB\_B33} polymorph: all six combinations of \texttt{FeB-19Wit} with \texttt{NdB-95Hal-LB} or \texttt{NdB-98Che} pass the edge check.
\item Silent introduction of the symbols \texttt{GBCCBB} / \texttt{GHCPBB}: no Nd--B parameter references them, and no \texttt{FeB-19Wit} symbol is redefined by the merge.
\end{itemize}

\subsection*{Conclusion of the diagnostic sequence}

The failures originate from a silent change in the sublattice model of the disordered bcc, fcc and hcp phases when \texttt{FeB-19Wit} (the only assessment that treats boron as purely interstitial in those phases) is merged with the two Nd--B assessments (\texttt{NdB-19Che}, \texttt{NdB-16Zho-corrected}) that treat boron as substitutional on the first sublattice of the same phases.
The resulting description remains within $6.2~\mathrm{J\,mol^{-1}}$ of the original Fe--B edge at a single test point.